\documentclass[aps,preprint,amsmath,amssymb]{revtex4}
\usepackage{graphicx}
\usepackage{color}

\def\g5{\gamma_{5}}
\def\ga{\gamma}

\def\be{\begin{eqnarray}}
\def\ed{\end{eqnarray}}

\def\non{\nonumber}

\begin{document}
\title{ \bf
New CP phase of $B_s-\bar B_s$ mixing on ${\rm T}$ violation in
$B_{d(s)}\to K^*(\phi) \ell^{+} \ell^{-}$ }
\date{\today}

\author{\bf  Chuan-Hung Chen$^{1,2}$\footnote{Email:
physchen@mail.ncku.edu.tw}, Chao-Qiang Geng$^{3}$\footnote{Email:
geng@phys.nthu.edu.tw} and Lin
Li$^{1}$\footnote{Email:lilin@phys.sinica.edu.tw}
 }

\affiliation{ $^{1}$Department of Physics, National Cheng-Kung
University, Tainan 701, Taiwan \\
$^{2}$National Center for Theoretical Sciences, Hsinchu 300, Taiwan\\
$^{3}$Department of Physics, National Tsing-Hua University, Hsinchu
300, Taiwan   }

\begin{abstract}
A large CP-violating phase uncovered recently by CDF and D$\O$
collaborations in the time-dependent CP asymmetry (CPA) of the
$B_s\to J/\Psi \phi$ decay clearly indicates that a
non-Cabibbo-Kobayashi-Maskawa (CKM) phase has to be brought into the
$b\to s$ transition. We find that the model with $SU(2)_L$ singlet
exotic quarks can not only provide the new phase induced from the
Z-mediated flavor changing neutral current (FCNC) at  tree level,
but also strongly relate the $B_s-\bar B_s$ mixing, $B_q\to V_q
\ell^{+} \ell^{-}$ ($V_{d[s]}=K^*[\phi])$ and $B_s\to \mu^{+}
\mu^{-}$ together. In particular, we show that the new CP phase can
be unambiguously exposed by the large statistical significances of
$T$ violating observables in $B_q\to V_q \ell^{+} \ell^{-}$, while
the branching ratio of $B_s\to \mu^{+} \mu^{-}$ can be enhanced to
be $O(10^{-8})$.
\end{abstract}
 \maketitle

CP violation (CPV) has been one of the most mysterious phenomena in high
energy physics since it was discovered
in the K system
\cite{CP}.
At B factories, BABAR and BELLE have observed
both the mixing-induced time-dependent CP asymmetry (CPA) in the $B_d$
oscillation through the golden  mode of $B\to J/\Psi K_S$
and the direct CPAs in exclusive $B\to \pi \pi$ and $B\to \pi K$
decays \cite{HFAG}, where the former is dictated by the $b\to d$
transition while the latter $b\to s$. Although three generations of
the standard model (SM) can provide a unique CP violating phase in
the Cabibbo-Kobayashi-Maskawa (CKM) matrix \cite{CKM} to interpret
the observed CPAs, it does not provide a solution to understand the
matter-antimatter asymmetry in the Universe and it does not stop
people to search for a new CPV source.

Nevertheless, it is very difficult to
unfold a new CPV phase in the direct CPAs of the nonleptonic exclusive
decays due to the inevitable large
uncertainty of nonperturbative QCD effects.
Hence, the best environment to look for the new phase is that
the QCD effects are less involved while the SM contributions are highly
suppressed. Now, the dawn to see the new effects could be in the
$B_s$ system. By the $B_s$ production in Tevatron Run II, besides CDF
and $D{\O}$ observations on the $B_s$ oscillation of $\Delta
m_s=17.77 \pm 0.10 \pm 0.07$ ps$^{-1}$ \cite{CDF} and $\Delta
m_s=18.56 \pm 0.87$ ps$^{-1}$ \cite{D0}, respectively, and the large
direct CPA of $0.39 \pm 0.17$ for $B_s\to K^{-} \pi^+$ \cite{BsCP},
an unexpected large CPV phase has been detected
in the mixing-induced CPA for $B_s\to J/\Psi \phi$.

To explain the new phase,
 we write the
transition matrix element for $\bar B_s\to B_s$ as
 \be
M^{s}_{12}&=& A^{\rm SM}_{12} e^{-2i\beta_s } + A^{\rm NP}_{12}
e^{2i(\theta^{\rm NP}_{s}-\beta_s)} \label{eq:ms12}
 \ed
where $\beta_s \equiv arg(-V_{ts} V^*_{tb}/V_{cs}V^*_{cb})$ is the
CPV phase in the SM and
$\theta^{NP}_{s}$ is the new CPV phase in
some extension of the SM. Here,
the convention has been chosen to be the same as that in Ref.~\cite{UTfit}. Due to
$ \Delta \Gamma_s \ll\Delta m_s$ in the B-system \cite{PDG06}, the
time-dependent CPA could be simplified to be
 \be
-S_{J/\Psi\phi}&\simeq& {\rm
Im}\left(\sqrt{\frac{M^{s^*}_{12}}{M^{s}_{12}}}\right)
=\sin(2\beta_s - \phi^{\rm NP}_{s})\,, \non\\
\phi^{NP}_{s}&=& \arctan\left( \frac{r\sin2\theta^{\rm
NP}_{s}}{1-r\cos2\theta^{\rm NP}_{s}} \right)
 \ed
with $r=A^{NP}_{12}/A^{\rm SM}_{12}$. By adopting Wolfenstein
parametrization \cite{Wolfenstein} of the CKM matrix up to
$O(\lambda^4)$, in which $V_{tb}=1-A^2\lambda^4/2$,
$V_{ts}=-A\lambda^2 + 1/2 A\lambda^4 \left( 1-2(\rho+i\eta)
\right)$, $V_{cb}=A\lambda^2$ and $V_{cs}=1-\lambda^2/2-1/8
\lambda^4(1+4A^2)$ \cite{Buras}, one can easily find
 \be
\beta_s \approx \lambda^2 \eta\approx 0.019\,,
 \ed
where $\lambda=0.2272$ and $\eta=0.359$ \cite{PDG06} have been  used. Clearly,
the mixing-induced CPA of $B_s$ in the SM is only few percent.
Astonishingly, the nonvanished CP phase measured
 by CDF
\cite{CDFCP} and D$\O$ \cite{D0CP} is
 \be
\phi_{s}=2\beta_s - \phi^{NP}_{s}= \left\{ \begin{array}{c}
                                     [0.24,\, 1.36]\cup [1.78,\, 2.82]\  (\textrm{CDF}) \\
                                    0.57 ^{+ 0.30+0.02}_{-0.24-0.07}
                                    \hspace{1.7cm} (\textrm{D$\O$})
                                   \end{array}
                                   \right.\label{eq:phis}
 \ed
at the $68\%$ confidence level (CL),
while the allowed range at
the $90\%$ CL by D$\O$ is given to be $\phi_{s}\in [-0.06,\, 1.20]$.
Recently, the UTfit collaboration has combined all available
information in the $B_s$ system and concluded that
the CPV phase of the  $B_s$ mixing amplitude deviates more than $3\sigma$
from the SM value
 \cite{UTfit}. To explain the large CPV phase in the $b\to s$ transition,
several new physics models have been proposed~\cite{models}.
In this paper, we explore the effects of the large phase
in the decays of
$B_q\to V_q \ell^{+} \ell^{-}$  ($V_{d(s)}=K^{*}(\phi)$),
corresponding to $b\to s \ell^+ \ell^-$ at the quark level. In particular, we will
show that the large CPV phase can be directly probed by measuring  T-odd observables
in the decays.

To comprehend the beauty of using T violation
to probe the
CPV phase, we briefly summarize the characters of CP-odd and
T-odd observables. In a decay process, the direct CPA or CP-odd
observable is defined by ${\cal A}_{CP}\equiv(\bar\Gamma -\Gamma)/(
\bar \Gamma+\Gamma)$, where $\Gamma$ ($\bar{\Gamma}$) is the partial
decay rate of the (CP-conjugate) process. As a result, ${\cal
A}_{CP}\propto\sin\theta_{w}\sin\theta_{st}$ with  $\theta_{w(st)}$ being the weak (strong) phase.
Clearly, to have a nonvanished CPA, both phases are needed.
The efficiency on the CPA is mainly dictated by the uncertain calculations of the
strong phase. Another way to probe the CPV phase is through the
spin-momentum triple correlation, such as  $\vec{s}\cdot (
\vec{p}_{i}\times \vec{p}_{j})$
\cite{Garisto-CGL,geng-BRD,CG_NPB636} for a three-body decay, where $\vec{s}$ is the spin
carried by one of outgoing particles and $\vec{p}_{i}$ and
$\vec{p}_{j}$ denote any two independent momenta. The
triple correlation is a T-odd observable since it changes
sign under the time reversal (${ \rm T}$)
transformation of $t\rightarrow -t$. We note that the T transformation defined here
is different from the real time-reversal transformation which also contains the
interchange of initial and final states.
By the CPT invariant theorem,
T violation (TV) implies
CPV. Therefore, the study of the T-odd observable can
help us to
understand the origin of CPV. Intriguingly,  the T-odd triple correlation
is proportional to $\sin(\theta_{w}+\theta_{st})$,
which indicates that the strong phase  is not necessary to achieve a nonzero
T-odd observable.
It has been shown in Ref. \cite{CG_NPB636}
that T-violating effects  in the exclusive $b\to s \ell^+ \ell^-$ processes are
sensitive to new physics with  small QCD uncertainties, which could provide
 a good place to directly observe the new
phase  revealed by CDF and D$\O$.

The transition amplitudes for $B_q\to V_q \ell^{+} \ell^{-}$
($\ell=e,\, \mu$) are given by \cite{CG_NPB636,CG_PRD66}
\begin{eqnarray}
{\cal M}_{V_q}^{(\lambda )} &=&-\frac{G_{F}\alpha \lambda_t}{2%
\sqrt{2}\pi }\left\{ {\cal M}_{1\mu }^{(\lambda )}L^{\mu }+{\cal
M}_{2\mu
}^{(\lambda )}L^{5\mu }\right\}\,,  \non \\
{\cal M}_{a\mu }^{(\lambda )} &=&if_{1}\varepsilon _{\mu \nu \alpha
\beta }\epsilon ^{*\nu }(\lambda )P^{\alpha }q^{\beta
}+f_{2}\epsilon _{\mu }^{*}(\lambda )+f_{3}\epsilon ^{*}\cdot
qP_{\mu }\,,   \non
\end{eqnarray}
where $\lambda_t=V_{tb}V_{ts}^{*}$, $L_{\mu}=\bar\ell \ga_{\mu}
\ell$, $L^5_{\mu}=\bar\ell \ga_{\mu}\ga_5 \ell$, $P=p_{B}+p_{V}$,
$q=p_{B}-p_{V}$,
$a=1\,(2)$ while $f_{i}=h_{i}$ $(g_{i})$
and
 \be
h_{1} &=&{C^{\rm eff}_9 V\over m_{B}+m_{V}}
+\frac{2m_{b}}{q^{2}}C_7T_{1}\,, \non \\
h_{2} &=&-\frac{1}{2}(m_{B}+m_{V})C^{\rm
eff}_9A_{1}-\frac{1}{2}\frac{2m_{b}}{q^{2}}
P\cdot q C_7 T_{2}\, ,  \non \\
h_{3} &=&{C^{\rm eff}_9 A_{2}\over m_{B}+m_{V}}+\frac{2m_{b}}{q^{2}}
C_7\Big(T_{2}(q^{2})+\frac{q^2}{P\cdot q}T_{3}\Big)\,,
\non \\
%
%
%
%
g_{i}&=&h_{i}|_{C^{\rm eff}_9\to C_{10}, C_{7}=0}\,,\ \ (i=1,2,3)\,.
  \label{eq:ampk1}
 \ed
Here, $m_{B}
(m_V)\equiv m_{B_q}(m_{V_q})$, $C^{\rm eff}_{9}$,  $C_7$ and $C_{10}$
are the Wilson coefficients \cite{BBL} and  the definitions of the
form factors in Eq. (\ref{eq:ampk1}) can be found in Ref.~\cite
{CG_NPB636}.
Furthermore, to obtain T-odd terms, the polarizations
of $V_q$ should be kept
in the averaged
squared-amplitude. To achieve the requirement, we have to consider
the decay chain $B_q\to V_q(\to P_1 P_2)\ell^{+} \ell^{-}$ in which
$P_1 P_2$ is $K\pi(KK)$ as $V_q=K^*(\phi)$. Consequently, the
differential decay rate associated with these
terms is given
by
 \be
&&\frac{d\Gamma }{d\cos \theta _{K}d\cos \theta _{\ell}d\phi
dq^{2}}=\frac{ 3\alpha^{2}G_{F}^{2}\left|\lambda_t \right|
^{2}\left| \vec{p}
\right| }{2^{14}\pi ^{6}m_{B}^{2}}  \nonumber \\
&&\times \{4\cos ^{2}\theta _{K}\sin ^{2}\theta
_{\ell}\sum_{i=1,2}|{\cal M}
_{i}^{0}|^{2}+\sin ^{2}\theta _{K}(1+\cos ^{2}\theta _{\ell})  \nonumber \\
&&\sum_{i=1,2}\left( |{\cal M}_{i}^{+}|^{2}+|{\cal
M}_{i}^{-}|^{2}\right)
-\sin 2\theta _{K}\sin 2\theta _{\ell}\sin \phi  \nonumber \\
&&\sum_{i=1,2}{Im}\left( {\cal M}_{i}^{+}-{\cal M}_{i}^{-}\right)
{\cal M}
_{i}^{0*}-2\sin ^{2}\theta _{K}\sin ^{2}\theta _{\ell}\sin 2\phi  \nonumber \\
&&\sum_{i=1,2}{Im}\left( {\cal M}_{i}^{+}{\cal M}_{i}^{-*}\right)
+2\sin
2\theta _{K}\sin \theta _{\ell}\sin \phi (Im{\cal M}_{1}^{0}  \nonumber \\
&&({\cal M}_{2}^{+*}+{\cal M}_{2}^{-*})-Im({\cal M}_{1}^{+}+{\cal
M}_{1}^{-}) {\cal M}_{2}^{0*})+\cdots ]\}\,, \label{eq:difrate}
 \ed
where $\theta_{\ell}(\theta_{K})$ is the polar angle of the lepton
(K-meson) in the $q^2$ ($V_q$) rest frame, $|\vec{p}
|=[((m_{B}^{2}+m_{V}^{2}-q^{2})/(2m_{B}))^{2}-m_{V}^{2}]^{1/2}$ and
$ {\cal M}_{i}^{0}$ and ${\cal M}_{i}^{\pm }$ denote the
longitudinal and transverse polarizations of $V_q$
with their explicit expressions
being
\begin{eqnarray}
{\cal M}_{a}^{0} &=&\sqrt{q^{2}}\left( \frac{E_{V}}{m_{V}}f_{2}+2
\sqrt{q^{2}}\frac{\left| \vec{p}_{V}\right| ^{2}}{m_{V}}f_{3}\right)
,\non \\
{\cal M}_{a}^{\pm } &=&\sqrt{q^{2}}\left( \pm 2\left|
\vec{p}_{V}\right| \sqrt{q^{2}}f_{1}+f_{2}\right) \,, \label{eq:Ma}
\end{eqnarray}
respectively.
To shorten the expressions, we have only presented the relevant pieces in
Eq.~(\ref{eq:difrate}), where the three imaginary terms denote the
T-violating effects. The whole expression for differential
decay rate could refer to Refs.~\cite{CG_NPB636,BHP}.

 From Eqs.~(\ref{eq:ampk1}), (\ref{eq:difrate}) and
(\ref{eq:Ma}),
it is easy to show that the large contributions to
the T-violating effects
arise from $Im{\cal M}_{1}^{0} ({\cal M}_{2}^{+*}+{\cal
M}_{2}^{-*})-Im({\cal M}_{1}^{+}+{\cal M}_{1}^{-}) {\cal
M}_{2}^{0*}$.
To explore the effects, we examine the T-odd observable, defined
by $\left\langle {\cal O}_T\right\rangle =\int {\cal O}_T d\Gamma $
where ${\cal O}_{T}$ is a T-odd five-momentum correlation, given by
\begin{equation}
{\cal O}_{T}=\frac{ \vec{p}_{B}\cdot \vec{p}_{K}}{\left| \vec{p}
_{B}\right| \left| \vec{p}_{K}\right|}\frac{ \vec{p} _{B}\cdot
(\vec{p}_{K}\times \vec{p}_{\ell^{+}}) }{\left| \vec{p}
_{B}\right|\left| \vec{p}_{K}\right|\omega_{\ell^{+}} } \,
\end{equation}
with $\omega_{\ell^{+}}=q\cdot p_{\ell^{+}}/\sqrt{q^2}$. In the
$V_q$ rest frame, we note that ${\cal O}_T=\cos \theta _{K}\sin
\theta _{K}\sin \theta _{\ell}\sin \phi $. To signal the nonvanished
CPV phase, we employ the statistical significance of the observable,
defined by
\begin{equation}
\varepsilon_T (q^{2})={\frac{\int {\cal O}_Td\Gamma }{\sqrt{(\int
d\Gamma )(\int {\cal O}^{2}_Td\Gamma )}}}\,.  \label{ss}
\end{equation}
Integrating all relevant angles in Eq. (\ref{ss}), we obtain
\begin{eqnarray}
\varepsilon_T (q^{2})&\simeq& \frac{0.76}{\sqrt{{\cal D}_{1}{\cal
D}_{2}}} [Im
{\cal M}_1^0({\cal M}_2^{+*}+{\cal M}_2^{-*})-  \nonumber \\
&& Im ( {\cal M}_1^++{\cal M}_1^-) {\cal M}_2^{0*}]\,,  \nonumber \\
{\cal D}_{a} &=&\sum_{i=1,2}\left[ \left| {\cal M}_{i}^{0}\right|
^{2}+\frac{1}{a}\left(\left| {\cal M}_{i}^{+}\right| ^{2}+\left|
{\cal M}_{i}^{-}\right| ^{2} \right)\right]. \,\
\label{eq:sig}
\end{eqnarray}
To observe the effect at $n\sigma $ level, the required number of B
mesons is $N_{B}=n^{2}/(Br\cdot \varepsilon^2_T)$.

We now use
the clue
of the current data to illustrate our model-independent analysis.
Although the principle of the minimal flavor violation
(MFV) \cite{MFV} could be as a dogma to rule the new source of CPV
\cite{MFVCP},
to focus on the criterion of the minimal extension of the SM,
we employ
the vector-like-quark model (VQM), in which the vector-like quarks
(VQs) are $SU(2)_L$ singlet exotic quarks, as the ones naturally
realized in $E_6$ models \cite{E_6}. Explicitly, we include
$SU(2)_L$ singlet VQs in the SM, where the right-handed component is
the same as the right-handed down-quark. Since the singlet quarks do
not couple to W-bosons directly, one of the fascinating characters
of the model is that the corresponding CKM matrix is not a unitary
matrix. By introducing flavor mixing matrices to diagonalize the
$4\times 4$ down-type quark mass matrix, we will display that the
model interestingly leads to Z-mediated flavor changing neutral
currents (FCNCs) at tree level \cite{ZFCNC}, which clearly have
significant impacts on the $B_s-\bar{B}_s$ mixing, $B_q\to V_q
\ell^{+} \ell^{-}$ and $B_s\to \mu^{+} \mu^{-}$ processes. We note
that the non-unitary CKM matrix could result in new contributions to
the processes from box and penguin diagrams.
However, to simply illustrate the new phase, only
the Z-mediated effects at tree level are considered here, while those from
the box and penguin diagrams could be referred to
Ref.~\cite{Aguilar-Saavedra}.

In the mass eigenstates, the coupling of  the $Z$-boson to fermions
 is written by
 \begin{eqnarray}
{\cal L}_{Z}&=& - \frac{gc^{f}_{L}}{2\cos\theta_W} \bar F \ga^{\mu}
\left(V^{L}_{F} X_{F} V^{L^\dagger}_{F}\right) P_{L}
  F Z_{\mu} \,,\non\\
X_{Q}&=& \left[
            \begin{array}{ccc}
                \begin{array}{c}
                  \openone_{3\times 3} \\
                \end{array}
              & | & {\bf 0}_{3\times 1} \\
             -\ -\ - & - &\  - \\
                              {\bf 0}_{1\times 3} & |& \xi_Q \\
            \end{array}
          \right]\,, \ \ \ X_{\ell}= \openone_{3\times 3}\,,
 \end{eqnarray}
where $g$ is the coupling constant of $SU(2)_{L}$, $\theta_W$ is the
Weinberg's angle, $P_{R(L)}=(1\pm\ga_5)/2$,
 $F^{T}=(q_1,q_2,q_3,q_4)$ and $(e,\mu,\tau)$
represent  quarks (Qs) and leptons $(\ell s)$, $c^{f}_{L}$ is
defined as $c^{f}_{L} = c^{f}_{V} +c^{f}_{A}$ with
 \be
 c^{f}_V &=& T^3_f -2\sin^2\theta_W Q_{f}\,,\ \ \ c^{f}_{A}=
 T^3_f
  \ed
in which $T^3_f$ and $Q_f$ are the third component of  the weak
isospin and the electric charge of the particle, respectively, and
$\xi_{f}=-2\sin^2\theta_W Q_f/c^{f}_{L}$. Here, as the quantum
number of the new right-handed VQ
is the same as the right-handed
down-quark, the tree FNCNs only occur in the left-handed quarks.
Accordingly, the interaction for $b$-$s$-$Z$ is given by
 \be
 {\cal L}_{b\to s}=\frac{g c^{D}_{L}\lambda_{23} }{2\cos\theta_W} \bar s \ga^{\mu}
  P_L b Z_{\mu} + H.c.
 \label{eq:bsint}
 \ed
with $\lambda_{23}=(1-\xi_D)(V^{L}_D)_{24}
(V^{L}_D)^*_{34}\equiv|\lambda_{23}|\exp[i(\theta^{\rm
NP}_{s}-\beta_s)]$. By Eqs.~(\ref{eq:ms12}) and (\ref{eq:bsint}),
the transition matrix element for the $\Delta B=2$ process is
obtained as
 \be
A^{\rm NP}_{12}&=& \frac{G_F \left(\lambda_{23}\right)^2}{3\sqrt{2}}
m_{B_s} f^2_{B_s} \hat{B_s}\,.
 \ed
As a result, the $B_s-\bar B_s$ mixing in the VQM is
 \be
 \label{Bs}
\Delta m_s=\Delta m^{\rm SM}_{s}\left( 1+2r\cos\theta^{\rm NP}_{s} +
|r|^2\right)^{1/2}\,.
 \ed
 From the above equation, it is clear that a large new CPV phase can have
a significant influence on the $B_s-\bar B_s$ mixing.
By combining the SM and Z-mediated FCNC, the
branching ratio (BR) of $B_s\to \mu^{+} \mu^{-}$ is found to be
 \be
{\cal B}_{\ell^+ \ell^-}=\tau_{B_s}\frac{G^2_F}{16\pi}
m_{B_s}f^2_{B_s}m^2_{\ell} \left(
1-\frac{4m^2_{\ell}}{m^2_{B_s}}\right)^{1/2}\left(|\Re|^2 + |\Im|^2
\right),\non
\ed
 \be
\Re &=& -\frac{|\lambda_t|\alpha}{\pi\sin^2\theta_W} Y(x_t)
+|\lambda_{23}| c^{D}_{L}\cos\theta^{\rm NP}_s,
\non\\
\Im &=& |\lambda_{23}|c^{D}_{L} \sin\theta^{\rm NP}_s.
 \ed
We can also  obtain the effects of the
Z-mediated FCNCs on $B_q\to V_q \ell^{+} \ell^{-}$ decays by
utilizing the replacement:
 \be
C^{\rm eff}_{9} [V,A_{1(2)}]&\to& \left(C^{\rm eff}_9 +
\frac{4\lambda_{23}}{\alpha \lambda_t} c^{D}_{L} c^{\ell}_{V}
\right)[V,A_{1(2)}],\non\\
 C_{10} [V,A_{1(2)}]&\to& \left(C_{10} - \frac{4\lambda_{23}}{\alpha
\lambda_t} c^{D}_{L} c^{\ell}_{A} \right)[V,A_{1(2)}].
 \ed
We note that in the following numerical analysis we will concentrate
on $B_d\to K^* \ell^{+} \ell^{-}$ as they have already  been
observed. However, the same discussions can be easily applied to $B_s\to
\phi \ell^{+} \ell^{-}$.

\begin{figure}[bpth]
\includegraphics*[width=3.3 in]{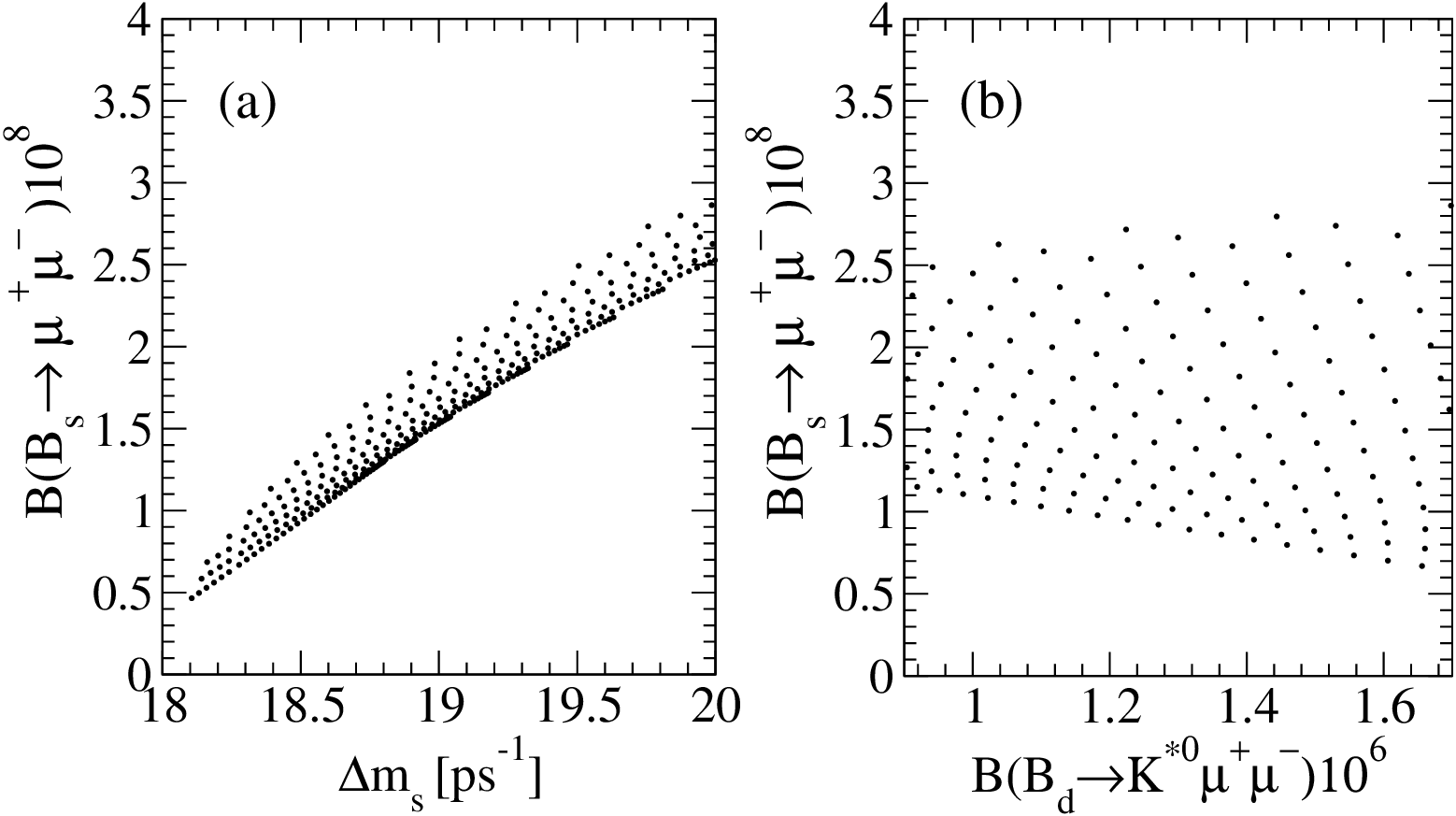}
\caption{(a)[(b)]
 ${\cal B}(B_s\to \mu^{+}\mu^{-})$ versus $\Delta m_s$
[${\cal B}(B_d\to K^{*0}\ell^{+} \ell^{-})$].}
 \label{fig:brs}
\end{figure}

\begin{figure}[bpth]
\includegraphics*[width=3.3 in]{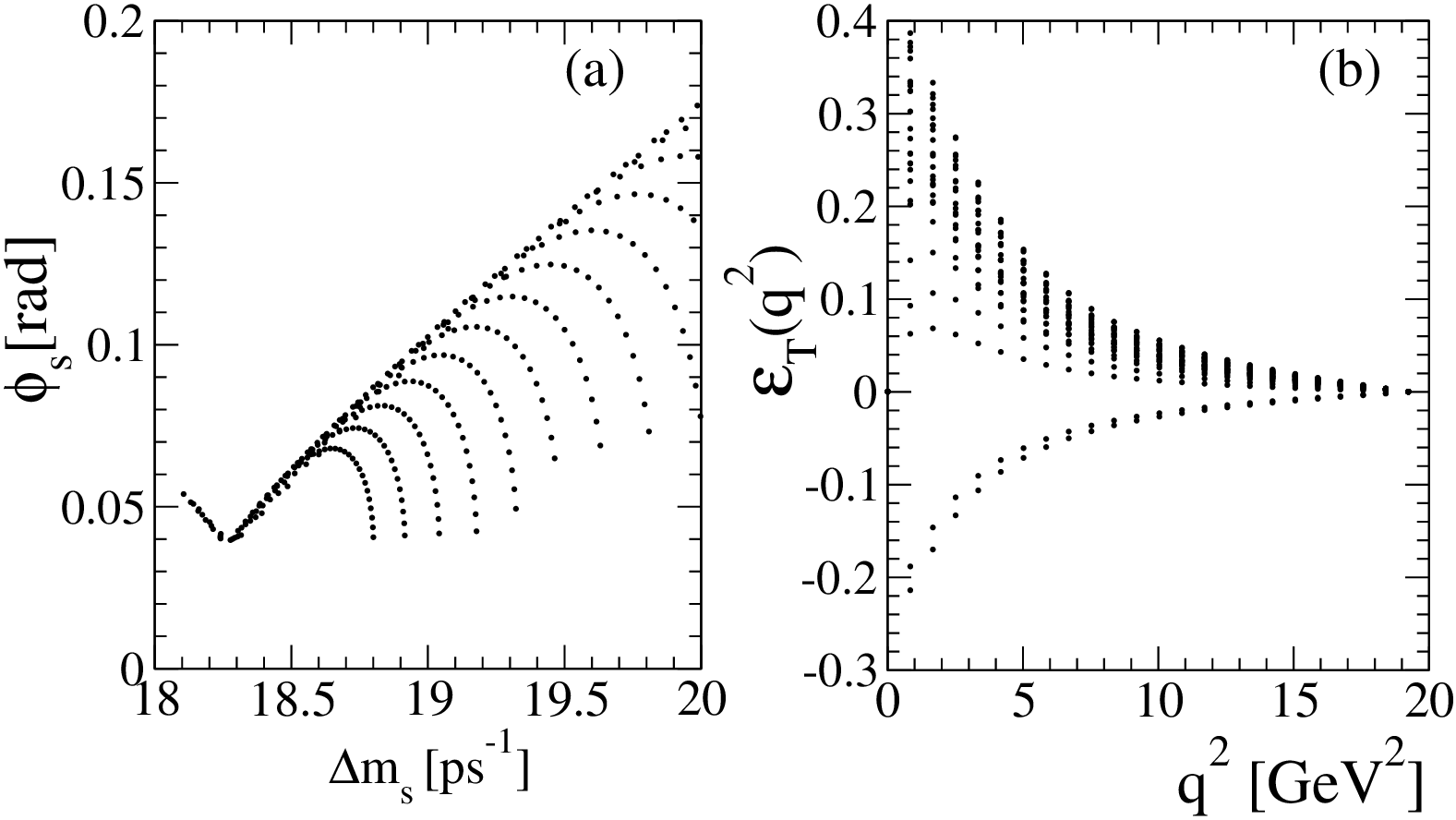}
\caption{(a) Correlation of $\phi_s=2\beta_s -\phi^{\rm NP}_{s}$ and
$\Delta m_s$. (b)
Statistical significance $\varepsilon_{T}$ of
$B_d\to K^{*0}\mu^{+} \mu^{-}$ as a function of $q^2$.}
 \label{fig:phis-todd}
\end{figure}

In the Z-mediated $b\to s$ transition, the magnitude
$|\lambda_{23}|$ and the phase $\theta^{\rm NP}_{s}$ can be
determined by the observed $B_s$ mixing and BR of $B_d\to
K^{*0}\mu^{+} \mu^{-}$. We adopt the average value of $\Delta
m_s=18.17\pm 0.86$ ps$^{-1}$ \cite{CDF,D0} and ${\cal B}(B_d\to
K^{*0}\mu^{+} \mu^{-})=(1.22^{+0.38}_{-0.32})\times 10^{-6}$
\cite{PDG06} as the inputs, while the SM results are taken to be
$\Delta m^{\rm SM}_s=19.3\pm 6.7$ ps$^{-1}$ \cite{LN}, ${\cal
B}(B_d\to K^{*0}\mu^{+} \mu^{-})_{\rm SM}=1.3\times 10^{-6}$
\cite{CGA21} and ${\cal B}(B_s\to \mu^{+} \mu^{-})_{\rm SM}\approx
0.33\times 10^{-8}$ with $V_{ts}=-0.04$ and $f_{B_s}=0.23$ GeV.
In order to reveal
the strong correlations among $\Delta m_s$,
${\cal B}(B_d\to K^{*0}\ell^{+} \ell^{-})$ and ${\cal B}(B_s\to
\mu^{+} \mu^{-})$ influenced by the same parameters, we present
${\cal B}(B_s\to \mu^{+} \mu^{-})$ versus $\Delta m_s$ [${\cal
B}(B_d\to K^{*0}\ell^{+} \ell^{-})$] in Fig.~\ref{fig:brs}(a)[(b)].
From the figures, we see clearly that the Z-mediated effects could enhance
the BR of $B_s\to \mu^{+} \mu^{-}$ to be $O(10^{-8})$,
which is close to the
current upper limit of $ 4.7\times 10^{-8}$ \cite{CDFBsmumu}.
With the constrained values of $\lambda_{23}$ and $\theta^{\rm
NP}_{s}$,   in
Fig.~\ref{fig:phis-todd}(a) we show the allowed $\phi_s$ given by Eq.~(\ref{eq:phis}).
It is wroth mentioning that ${\cal
B}(B_s\to K^{*0} \ell^{+} \ell^{-})$ excludes
$\phi_s$
larger than 0.2 rad. Furthermore, we present the
contributions of the new CP violating source to the T-odd observable
of Eq.~(\ref{eq:sig}) in Fig.~\ref{fig:phis-todd}(b). Intriguingly,
the new phase can lead to a large statistical significance of the T-odd
observable in $B_q\to V_q\ell^{+} \ell^{-}$.

In summary, motivated by the large CP phase found by CDF and D$\O$
in the $B_s-\bar B_s $ mixing, we have investigated the $SU(2)_L$
singlet VQM. This model can not only provide the large phase through
the Z-mediated FCNCs at  tree level, but also strongly relate
$\Delta m_s$, $B_q\to V_q \ell^{+} \ell^{-}$ and $B_s\to \mu^{+}
\mu^{-}$ processes. In particular, we have shown that the new CP
phase can be unambiguously exposed by the large statistical
significances of the T-violating observables in $B_q\to V_q \ell^{+}
\ell^{-}$ ($V_{q}=K^*,\, \phi)$. In addition, we have found that
${\cal B}(B_s\to \mu^{+}\mu^{-})$ can be enhanced as large as
$O(10^{-8})$. Finally, we remark that the T-violating effects in
$B_q\to V_q \ell^{+}\ell^{-}$ as well as the result on ${\cal
B}(B_s\to \mu^{+}\mu^{-})$ are accessible at future super-B
factories, such as the SuperKEKB
 \cite{SuperKEKB}
 and LHCb \cite{LHCb}.
For example,
4400 events/year for
$B\to K^* \ell^{+} \ell^{-}$ decays
will be produced at the LHCb, corresponding to the
accuracy of the T-odd observable being around
percent level.

\begin{acknowledgments}
 This work is supported in part by
the National Science Council of R.O.C. under Grant \#s: NSC
97-2112-M-006 -001-MY3 and NSC-95-2112-M-007-059-MY3.

\end{acknowledgments}

\end{document}